\documentstyle[aps,twocolumn,epsf,psfig]{revtex}

\begin{document}
\newtheorem{theorem}{Theorem}
\newtheorem{acknowledgement}[theorem]{Acknowledgement}
\newtheorem{algorithm}[theorem]{Algorithm}
\newtheorem{axiom}[theorem]{Axiom}
\newtheorem{claim}[theorem]{Claim}
\newtheorem{conclusion}[theorem]{Conclusion}
\newtheorem{condition}[theorem]{Condition}
\newtheorem{conjecture}[theorem]{Conjecture}
\newtheorem{corollary}[theorem]{Corollary}
\newtheorem{criterion}[theorem]{Criterion}
\newtheorem{definition}[theorem]{Definition}
\newtheorem{example}[theorem]{Example}
\newtheorem{exercise}[theorem]{Exercise}
\newtheorem{lemma}[theorem]{Lemma}
\newtheorem{notation}[theorem]{Notation}
\newtheorem{problem}[theorem]{Problem}
\newtheorem{proposition}[theorem]{Proposition}
\newtheorem{remark}[theorem]{Remark}
\newtheorem{solution}[theorem]{Solution}
\newtheorem{summary}[theorem]{Summary}
\def\r{{\bf{r}}}
\def\i{{\bf{i}}}
\def\j{{\bf{j}}}
\def\m{{\bf{m}}}
\def\k{{\bf{k}}}
\def\kt{\tilde{\k}}
\def\K{{\bf{K}}}
\def\P{{\bf{P}}}
\def\q{{\bf{q}}}
\def\Q{{\bf{Q}}}
\def\p{{\bf{p}}}
\def\x{{\bf{x}}}
\def\X{{\bf{X}}}
\def\Y{{\bf{Y}}}
\def\F{{\bf{F}}}
\def\G{{\bf{G}}}
\def\M{{\bf{M}}}
\def\V{\cal V}
\def\tchi{\tilde{\chi}}
\def\tk{\tilde{\bf{k}}}
\def\tK{\tilde{\bf{K}}}
\def\tq{\tilde{\bf{q}}}
\def\tQ{\tilde{\bf{Q}}}
\def\si{\sigma}
\def\ep{\epsilon}
\def\al{\alpha}
\def\be{\beta}
\def\ep{\epsilon}
\def\up{\uparrow}
\def\de{\delta}
\def\De{\Delta}
\def\up{\uparrow}
\def\dwn{\downarrow}
\def\ksi{\xi}
\def\etha{\eta}
\def\product{\prod}
\def\goto{\rightarrow}
\def\switch{\leftrightarrow}

\title{Absence of a Slater Transition in The Two-Dimensional Hubbard Model} 
\author{S. Moukouri and M. Jarrell }
\address{Department of Physics, University of Cincinnati, 
Cincinnati, OH 45221}
\date{\today }
\maketitle

\begin{abstract}
We present well-controlled results on
 the metal to insulator transition (MIT) within the paramagnetic
solution of the dynamical cluster approximation (DCA) in the two-dimensional
Hubbard model at half-filling. In the strong coupling regime, a local 
picture describes the properties of the model; there is a large charge gap 
$\Delta \approx U$. In the weak-coupling regime, we find a symbiosis of 
short-range antiferromagnetic correlations and moment formation cause a gap 
to open at finite temperature as in one dimension.  Hence, this excludes the 
mechanism of the MIT proposed by Slater long ago.  

\end{abstract}


\paragraph*{Introduction} In this letter we report the study of  
the metal-insulator transition (MIT) and its relation to antiferromagnetism 
(AFM) in the two-dimensional (2D) Hubbard model at half-filling. Results from 
numerical simulations \cite{hirsch1,white1} have convincingly shown 
that the ground state of the model is an AF insulator, with the N\'eel 
temperature constrained to be $T_{N}=0$ by the Mermin-Wagner theorem. 
However, the nature of the MIT is less clear, as there are two conflicting 
opinions concerning the MIT and its relation to AFM. The {\em{first 
opinion}} \cite{white2,nozieres,schrieffer} is that the strong and weak 
coupling MI transitions are very different.  When the local Coulomb 
repulsion parameter $U$ is larger than the non-interacting band width $W$,  
the ground state is an insulator with a large charge gap $\Delta \approx U$.  
The MIT occurs well before the onset of magnetism at the temperature 
$T_g \approx U$, and the spin and charge degrees of freedom are decoupled. 
The superexchange interaction couples the spins with the exchange constant 
$\left |J\right | \approx 4t^2/U$, and the spins govern the low-energy 
physics.  This type of MIT, which is purely due to local correlations, 
is called a Mott transition. In weak coupling, a spin density wave (SDW) 
instability develops at $T=0$ because of the nesting of the Fermi surface. 
The MIT is the direct consequence of the Brillouin zone folding generated 
by magnetic ordering. This type of MIT will be referred to as a Slater 
transition. It is believed that this regime can be well described by the 
usual many-body weak-coupling approaches.  The {\em{second opinion}}, is 
due to Anderson\cite{anderson} who has argued that the strong-coupling 
behavior presented above occurs for both strong and weak coupling so that 
a Mott gap is present for all $U > 0$ as in one dimension.  As the 
temperature falls, local moments develop first because of the MIT and 
then they order so that AFM is the consequence of the MIT, not the 
converse. Because of this strong interaction, there is no adiabatic 
continuity between the non-interacting and interacting eigenstates. The 
conventional renormalizable many-body perturbation theory cannot describe 
this physics.   

The nature of the MIT has been addressed in the Hubbard model in infinite
dimensions\cite{jarrell,georges,rozenberg} by applying the dynamical 
mean-field approximation (DMFA)\cite{dmfa}. The unfrustrated model is an 
example of the first scenario presented above with a Slater AF insulator 
in weak coupling, and a Mott MIT at strong coupling. The one-dimensional
model is an example of the second scenario, since there is a gap in the 
charge excitation spectrum for all non-zero $U$\cite{lieb}, but no long 
range magnetic order. 

In 2D, the Mermin-Wagner theorem precludes any AFM transition at any finite
temperature. Hence, if there is a MIT  at finite $T$, it cannot be attributed 
to the Slater mechanism.  Finite size simulations (FSS) employing the quantum 
Monte Carlo (QMC) technique have been applied to study this model quite 
extensively during the last decade\cite{hirsch1,white1,white2,moukouri1}.
But the difficulty with FSS is that at low temperatures, the correlation 
length is greater than the lattice size. Thus the effects of correlations 
are overestimated for smaller clusters because they are artificially 
closer to criticality than a system in the thermodynamic limit. This 
tendency is only reduced by increasing the cluster size, which moves the 
system in the direction of the thermodynamic limit. The spurious AFM gap 
that opens at finite $T$ in the FSS renders the disentanglement of the MI
and the AF transitions practically impossible.  The situation is radically 
different in the dynamical cluster approximation (DCA), which is an 
extension of the DMFA. In the DCA, the system is already in the 
thermodynamic limit; however, the DCA restricts correlations to within the 
cluster length.  As the cluster size increases, longer range correlations 
are progressively included.  The FSS and DCA results are complimentary in 
that the DCA (FSS) always underestimates (overestimates) the effects of 
non-local correlations\cite{huscroft,moukouri2}. Thus, the DCA will always 
systematically underestimate the gap formation inherent in the Anderson 
picture of the MIT. Therefore, if a gap is found in the DCA for a finite 
cluster, it will persist in the thermodynamic limit. 

Using the DCA, we show that a MIT occurs in the 2D Hubbard model at 
half-filling and  finite temperature at both weak and strong coupling 
regimes. This result is to be contrasted with the weak-coupling approaches 
which predict that a gap will only exist at $T=0$ as a consequence of AFM. 
We argue that at weak couplings the symbiosis of short-range AFM correlations 
and local moment formation is the key mechanism for the opening of a gap. 
The local moments forms at a relatively high $T$ leading to a short-range 
AFM order, that in turn enhances the moments which enhance the AF order, etc.
This leads to the destruction of the Fermi liquid state found in the DMFA 
studies \cite{GKKR}.  We discuss the validity of our conclusions drawn from 
finite clusters at the thermodynamic limit.

\paragraph*{Method } A detailed discussion of the DCA formalism was given in 
previous publications \cite{DCA_hettler,DCA_maier1,moukouri2}.  The compact 
part of the free energy is coarse-grained in reciprocal space, projecting 
the problem onto a finite-sized cluster of $N_c$ points embedded in a 
self-consistently determined host. When $N_c=1$, the DCA recovers the DMFA, 
and when  $N_c \rightarrow \infty$, it becomes exact. The cluster problem 
is solved with a generalized version of the Hirsch-Fye QMC algorithm\cite{fye},
and the spectra are analytically continued with the maximum entropy 
method\cite{JARRELLandGUB}.  We use the DCA to study the single band Hubbard 
model in a two-dimensional square lattice at half-filling. The model is 
characterized by $W$ and $U$. We set $W=2$ and we vary $U$ from $U=W/4$ to
$U=2W$.

\begin{figure}[ht]
\leavevmode\centering\psfig{file=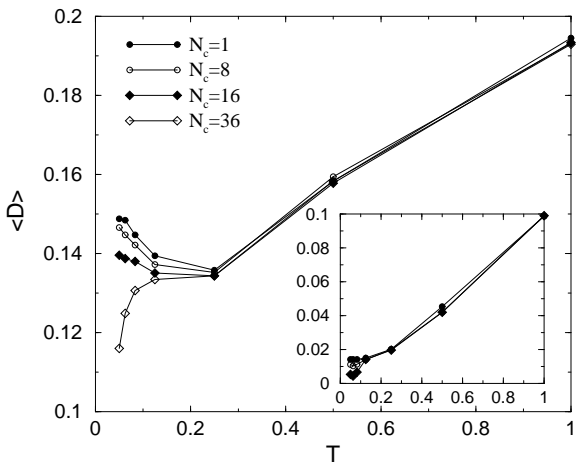,width=3. in}
\caption{The double occupancy D for $U=1$ and $U=3$ (inset) for various 
$N_c$.}
\label{doubc3}
\end{figure}

\paragraph*{Results}
Jarrell \cite{jarrell}, Georges and Krauth \cite{georges} and Rozenberg
et al. \cite{rozenberg} originally obtained a numerically exact solution 
from the DMFA in the limit of infinite dimensions using QMC. The analysis 
of these mean-field equations revealed the existence of a transition between 
a paramagnetic state and an AFM state.  The paramagnetic solution  was 
shown to be a Fermi liquid for small values of the interaction $U$ and an
insulator for large $U$\cite{remark}.  Georges and Krauth identified this 
transition by observing the behavior of the double occupancy which is
given by  
$\langle D \rangle =\langle n_{\uparrow}n_{\downarrow}
\rangle$, where $n_{\sigma}$ is the
density of spin $\sigma$, and of the single-particle 
Green's function $G(\tau)$. $\langle D \rangle(T)$ displays a minimum in the 
Fermi liquid regime. But when the system is an insulator for $U > U_c$, 
$\langle D \rangle $ always increases with T. We observe the same 
qualitative behavior as in\cite{georges} when $N_c=1$.

We will perform the same kind of analysis to study the effect of short-range 
AFM correlations that are absent in the DMFA.  The behavior of 
$\langle D \rangle $ as function of $N_c$ depends strongly upon $U$. In 
the strong coupling regime, $\langle D \rangle (T)$ is essentially independent 
of $N_c$ (c.f inset to Fig.~\ref{doubc3}). This was expected, since for large 
$U$, local fluctuations dominate; they are already captured in $N_c=1$. In 
the weak coupling regime, the minimum found for $N_c=1$ flattens progressively 
as $N_c$ increases from $8$ to $16$. At  $N_c=36$, there is a downturn
in $\langle D \rangle $ at low $T$. By localizing the moments at low 
temperatures, the system gains free energy by taking advantage of the short 
range magnetic order.  For the value $U=1$, we are already deep in the 
weak-coupling regime of the model. Hence, we expect this behavior to be 
generic in this region.  This finding is similar to a recent finite system 
QMC simulation \cite{paiva}.   


\begin{figure}[ht]
\leavevmode\centering\psfig{file=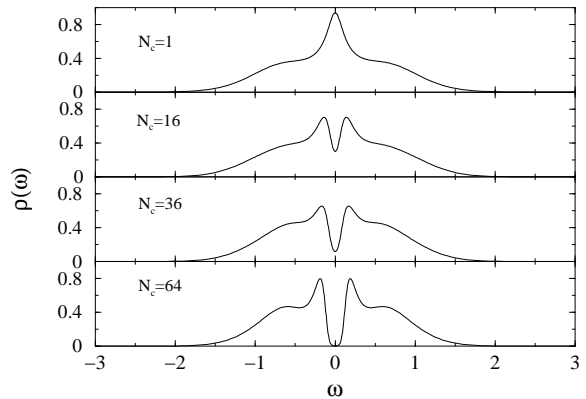,width=3. in}
\caption{  The DOS $\rho(\omega)$ at $\beta=32$ and $U=1$.}
\label{doscu1}
\end{figure}
The density of states (DOS) $\rho(\omega)$ 
(shown in Fig.~\ref{doscu1}) confirms the destruction 
of the Fermi liquid quasi-particle peak by short-range AFM correlations.
 With increasing $N_c$, the gap opens fully, and the Hubbard side bands
become more pronounced, consistent with the suppression of $\langle D \rangle $ 
shown in Fig.~\ref{doubc3}.  Since we work within the paramagnetic equations, 
the opening of the gap in the weak-coupling regime cannot be attributed to 
the Slater mechanism. 

\begin{figure}[ht]
\leavevmode\centering\psfig{file=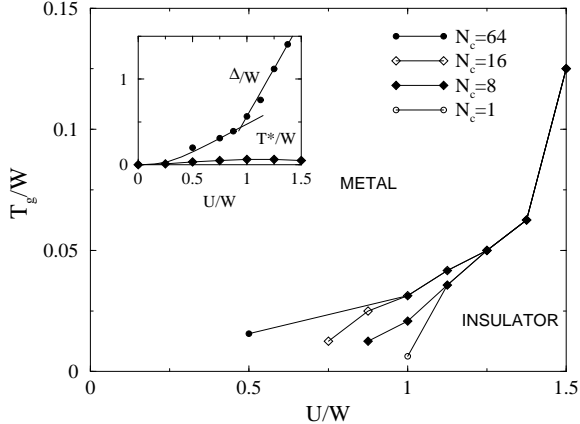,width=3. in}
\caption{  The temperature $T_g$ the gap opens in $\rho(\omega)$ 
$N_c=1, 8, 16, 64$ as function of $U$. In the inset the mean-field
 transition temperature $T^*$ and the gap $\Delta$ as
function of $U$, the linear ($U > W$) and exponential ($U < W$)
fits for $\Delta$ serve as guide to eyes.}
\label{phase}
\end{figure}

In Fig.~\ref{phase} we show a tentative phase diagram obtained with 
$N_c=1,8,16, 64$.  The boundary line between the metallic and the 
insulating phases is defined as the point where a true gap opens to 
the system within our numerical accuracy; the criterion for the gap 
opening is roughly  $\rho(\omega=0) < 1 \times 10^{-2}$.  Hence for 
temperatures just above this line, $\rho(\omega)$ displays a pseudogap. 
As expected, for $U > W$, the boundary line is almost independent of 
the cluster size. For $U < W$ however, the $N_c=1$ line ends at 
$U=U_c \approx W$. For $U < U_c$, a Fermi liquid is stable at low $T$. 
For larger clusters, we don't find a quasiparticle peak, except when 
$U\alt U_c$, and $U_c$ seems to approach zero as $N_c$ increases. This 
seems to indicate that {\it there is no weak-coupling fixed point in 
the two dimensional Hubbard model at half-filling}.  However, since an 
exponential behavior of the gap is expected in this limit, the numerical 
cost becomes too high at very small couplings; the smallest value of 
$U$ we were able to observe a pseudogap  is $U=0.5$. For instance, it 
was necessary that $N_c > 36$ in order to see a gap at $U =1$ (as shown 
in Fig.~\ref{doscu1}). We believe that this gap is due to a synergism 
between short ranged order and moment formation as evidenced by the 
suppression of $\langle D \rangle $ (Fig.~\ref{doubc3}) and the enhancement 
of the Hubbard side bands (Fig.~\ref{doscu1}) seen when the gap opens.  
                       
It has been argued that the SDW picture might capture the
description of the insulating state  of the cuprates
in the weak to intermediate coupling regime \cite{bulut}.
We show in the inset to Fig.~\ref{phase} the gap $\Delta$ 
and the mean-field transition temperature, which to a very 
good approximation is also the pseudogap temperature $T^*$ 
characterized by a peak in the bulk susceptibility and the 
onset of the pseudogap in the DOS. $T^*$ defines the 
temperature where short-ranged AF correlations emerge.  If 
the gap were due only to these correlations, then we expect 
$\Delta \approx T^*$. For example, in the SDW approximation 
$\Delta/T^* = 1.76$\cite{fazekas}.  We find that $\Delta/T^*$ 
strongly deviates from this prediction.  In fact,  we find 
that $\Delta \gg T^* > T_g$ hence, SDW ordering is not 
the mechanism that leads to the gap formation.   

\begin{figure}[ht]
\leavevmode\centering\psfig{file=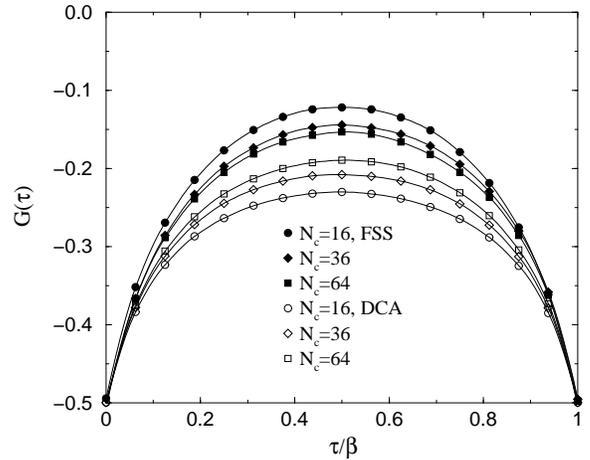,width=3. in}
\caption{The imaginary-time Green's function at the Fermi point $X=(\pi,0)$ 
calculated with finite size QMC (filled symbols) and with the DCA (empty 
symbols) when $U=1.1$, $\beta=16$.}
\label{gtau}
\end{figure}

We now show that the conclusions drawn from finite clusters are valid in 
the thermodynamic limit. For this, we exploit the complementarity between 
the DCA and FSS \cite{moukouri2,huscroft}.  In Fig.~\ref{gtau}, we show the 
imaginary-time Green's function $G(\tau)$ at the Fermi point $X=(\pi,0)$. 
This quantity has a more rapid decay  from it maximum at $G(\beta/2)$ 
when the effects of the correlations are stronger so that the gap is
more pronounced. In finite systems, the decay is sharper for smaller
lattices while in the DCA the opposite occurs. This behavior marks the
fundamental difference between the FSS and the DCA. At low temperatures,
in FSS, the correlation length is greater than the lattice size. Thus
the effect of correlations are overestimated for smaller clusters because 
these systems are artificially closer to criticality than a system in the 
thermodynamic limit. This tendency is reduced by increasing the cluster 
size. The situation is radically different in the DCA where the system is 
already in the thermodynamic limit.  The DCA restricts correlations to 
within the cluster length. As the cluster size increases, long range 
correlations are progressively included. Thus, the effects of the 
correlations increase with cluster size. Since the two techniques must
become identical in the thermodynamic limit, the $G(\tau)$ curve in 
the this limit is bracketed by the FSS and DCA curves.

\begin{figure}[ht]
\leavevmode\centering\psfig{file=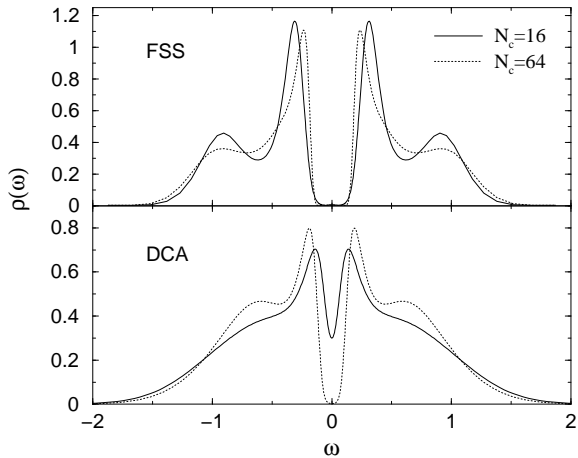,width=3. in}
\caption{  The DOS $\rho(\omega)$ from finite size QMC (top)
and from DCA (bottom) at $U=1$, $\beta=32$ and $N_c=16, 64$.}
\label{doscf}
\end{figure}

This complementarity is also seen in the spectra.  In the weak-coupling 
regime, the FSS and the DCA start from two different physical limits.  A 
finite system is always an insulator as soon as $T$ is less than the energy 
separation between the ground state and the first excited state. Hence finite 
systems always show gaps or pseudogaps if $T$ is sufficiently low. On the 
contrary, for $N_c=1$, the DCA is identical to the DMFA which at weak couplings 
yields a local Fermi metal. As the cluster size increases, one expects the two 
techniques, that become identical in the thermodynamic limit, to converge 
towards this limit from complementary directions. In the strong coupling 
regime, the DMFA has a gap too. But this it is always smaller than the one 
found in the FSS for the same set of parameters. Hence the convergence towards 
the thermodynamic limit in the DCA will be from smaller gaps to larger, while 
the converse occurs in FSS. The DOS shown in Fig.~\ref{doscf} supports these 
conclusions. The finite size gap in the FSS decreases when the cluster size 
goes from $N_c=16$ to $64$. In the DCA, there is a pseudogap for $N_c=16$ that 
turns into a true gap when the cluster size is increased to $64$.  Since by 
construction the DCA underestimates the gap, we can affirm that at this 
temperature, the gap exists in the thermodynamic limit.  Its actual value is 
bracketed by the FSS and the DCA.
 
\paragraph*{Conclusion} We have presented well-controlled
 results on the  MIT in the 2D Hubbard model obtained by
the DCA. We find that the double occupancy and the single-particle DOS show 
evidence of a gap for both small and large values of $U$ at finite $T$.
Since, a finite-temperature gap persists well into the weak coupling regime, 
and there is no long range order until $T=0$, the Slater mechanism is 
likely not responsible for the metal-insulator transition in the two 
dimensional Hubbard model. 
  Instead, the gap is due to a synergism between short ranged 
order and moment formation, as evidenced by the concomitant opening of the
gap and the suppression of the double occupancy.  Since the charge 
fluctuations are suppressed at finite $T$ by the MIT and the spins do not 
order until T=0, spin-charge separation also persists into the weak coupling 
regime.  As the DCA systematically underestimates the gap formation, these 
conclusions are valid in the thermodynamic limit.  The resulting phase 
diagram is consistent with Anderson's view that the effective Hamiltonian
for the 2D Hubbard model at half-filling  for all $U>0$ and $ \Delta \gg T$
is the 2D Heisenberg Hamiltonian\cite{anderson}.  Finally, our results put 
some constraints on theories of high temperature superconductivity. Since 
theories based on weak-coupling expansions are unable to describe these 
results, their prediction in the doped case are questionable.

\paragraph*{Acknowledgments }
Part of this work was performed during the visit of S.M. to the 
University of Southern California, we would like to thank  G.\ Bickers 
for the warm hospitality and for useful conversations.  We gratefully 
acknowledge discussions with G. Baskaran which motivated this work,  
we thank C. Huscroft for his help on parallel programming, and  
we thank M. Ma for helpful discussions.  This work 
was supported by the National Science Foundation grants DMR-0073308 and 
PHY94-07194 and the Ohio Supercomputing Center.


\begin{references}
\bibitem{remark}The authors of \cite{georges} and \cite{rozenberg} later
found hysteresis at very low temperatures from which they infer a first
order transition (see \cite{GKKR}). However, all of the calculations 
presented here occur at higher temperatures, and for $N_c \gg 1$, we do not 
observe a Kondo peak which was prerequisite for their scenario at the MIT.
\end{references}
\end{document}